# Autoresonance energy transfer versus localization in weakly coupled oscillators

Agnessa Kovaleva[1] and Leonid I. Manevitch[2]

[1] *Space Research Institute, Russian Academy of Sciences, Moscow 117997, Russia*
[2] *Institute of Chemical Physics, Russian Academy of Sciences, Moscow 119991, Russia*

**Abstract** – In this paper we investigate the distribution of energy between weakly coupled linear and nonlinear oscillators in a two-degree-of-freedom (2D) system. Two classes of problems are studied analytically and numerically: (1) a periodic force with constant frequency is applied to the nonlinear (Duffing) oscillator with slowly time-decreasing linear stiffness; (2) the time-independent nonlinear oscillator is excited by a force with slowly increasing frequency. In both cases, stiffness of the attached linear oscillator and linear coupling remain constant, and the system is initially engaged in resonance. This paper demonstrates that in the systems of the first type autoresonance (AR) occurs in both oscillators while in systems of the second type AR occurs only in the excited nonlinear oscillator but the coupled linear oscillator exhibits small bounded oscillations. Considering slow detuning, we obtain explicit asymptotic approximations for the amplitudes and the phases of oscillations close to exact (numerical) results.

**Introduction and motivation.** - Resonance energy transfer has been identified as an underlying mechanism for energy transport from a source to a receiver for a broad range of physical phenomena. Theoretical approaches, experimental evidence, and applications of this effect in different fields of physics and engineering have been reported and widely discussed (see, e.g., [1-3] and references therein). Special attention has been paid to resonance energy transfer and exchange between oscillators with constant parameters excited either by a periodic force or by an initial impulse. However, practical realization of the purposive energy transport even in these well-studied systems is often prevented by the fact that the non-stationary nonlinear problems of energy transport seldom yield explicit analytical solutions needed for understanding and modelling the transition phenomena.

In this work we analyze autoresonance (AR) energy transport between weakly coupled linear and nonlinear oscillators in a two-degree-of-freedom (2D) system. We recall that AR exploits the ability of a nonlinear oscillator to adjust its amplitude so that its nonlinear frequency matches its drive frequency. If the oscillator is initially in resonance, the synchronous change of these two frequencies maintains the resonance state with gradually increasing amplitude. This excitation scheme is relatively insensitive to both system characteristics as well as to the rate of the slow passage through resonance, thereby providing a way to achieve the designed energy level.

AR was first used in applications to particle acceleration and reported as "the phase stability principle" [4, 5]. Building on that works, a large number of theoretical, numerical and experimental results have been reported in the literature, see, e.g., [6-8] and references therein. The analysis was first concentrated on the basic single-degree of freedom model but then the developed methods and approaches were extended to two- or three-dimensional systems. Examples in this category are interactions of the plasma waves with laser beams [8-11], particle transport in a weak external field with slowly varying frequency [12, 13], control of diatomic molecules [14], etc.

Some particular results (e.g.,[9,10]) have demonstrated that external forcing with a slowly varying frequency applied to the pair of coupled nonlinear oscillators generates AR in both oscillators. In this work we demonstrate that this conclusion cannot be applied universally; furthermore, it does not hold for the pair of weakly coupled linear and nonlinear (Duffing) oscillators in the 2D system. This system represents, e.g., a simplified model of an asymmetric micro-oscillator cell [15]. We investigate two types of autoresonant problems for this system: (1) a periodic force with constant frequency acts on the Duffing oscillator with slowly time-decreasing linear stiffness; (2) the time-independent nonlinear oscillator is excited by a force with slowly increasing frequency. In both cases, stiffness of the linear oscillator and linear coupling remain constant, and the system is initially engaged in resonance.

It is important to note that the averaged response and conditions of the occurrence of AR in a single Duffing oscillator are equivalent for both types of excitation (see, e.g. [16, 17]). However, the behavior of each oscillator in the coupled system may drastically differ from the dynamics of a single oscillator. We demonstrate that periodic forcing with constant (resonant) frequency causes AR in both oscillators but the drive with the slowly-varying frequency gives rise to AR only in the excited nonlinear oscillator while the attached oscillator exhibits small bounded oscillations. Hence AR may serve as a method of creating a high-energy regime in the coupled linear oscillator only in the system of the first type, while in the system of the second type energy is localized on the excited nonlinear oscillator. To the best of authors' knowledge, this effect has not been discussed earlier.

[a] E-mail:agnessa_kovaleva@hotmail.com

The paper is organized as follows. In Sec. II the system of the first type is considered. It is shown that, under certain assumptions, the earlier obtained conditions of the occurrence of AR in a single oscillator [16, 17] may be extended to the 2D system. Given very small detuning rate, we obtain approximate solutions describing autoresonant phenomenon in both oscillators. The system of the second type is investigated in Sec. III. It is shown that AR in the nonlinear oscillator occurs along with small bounded oscillations of the attached oscillator. The difference in the dynamical behavior may be explained by different resonance properties of the systems. From the practical standpoint this means that AR as a method of formation of the required dynamics is effective only for the systems of the first type. Saturated oscillations at large detuning rate are investigated in Sec. IV.

**Energy transfer in the system of the first type.** – Consider the following model of two coupled oscillators:

$$m_1 \frac{d^2 u_1}{dt^2} + c_1 u_1 + c_{10}(u_1 - u_0) = 0,$$
$$m_0 \frac{d^2 u_0}{dt^2} + C(t)u + ku^3 + c_{10}(u_0 - u_1) = A\cos\omega t, \quad (1)$$

where $u_0$ and $u_1$ denote absolute displacements of the nonlinear and linear oscillators, respectively; $m_0$ and $m_1$ are their masses; $c_1$ and $k$ are the coefficients of linear stiffness and cubic nonlinearity; $c_{10}$ is the linear coupling coefficient; $C(t) = c_0 - (k_1 + k_2 t)$, $k_{1,2} > 0$; $A$ and $\omega$ are the amplitude and the frequency of the periodic force. The system is initially at rest, that is, $u_r = 0$, $v_r = du_r/dt = 0$ at $t = 0$, $r = 0, 1$.

Considering weak coupling, we introduce the small parameter of the system as $2\varepsilon = c_{10}/c_1 \ll 1$. Considering weak nonlinearity and taking into account resonance properties of the system, and then denote:

$$c_1/m_1 = c_0/m_0 = \omega^2, \; A = \varepsilon m\omega^2 F. \quad (2)$$
$$\tau_0 = \omega t, \; \tau_1 = \varepsilon \tau_0, \; k_1/c_0 = 2\varepsilon s,$$
$$k_2/c_0 = 2\varepsilon^2 b\omega, \; k/c_0 = 8\varepsilon\alpha, \; c_{10}/c_1 = 2\varepsilon\lambda_1, \; c_{10}/c_0 = 2\varepsilon\lambda_0.$$

In these notations, the equations of motion take the form:

$$\frac{d^2 u_1}{d\tau_0^2} + u_1 + 2\varepsilon\lambda_1(u_1 - u_0) = 0, \quad (3)$$
$$\frac{d^2 u_0}{d\tau_0^2} + (1 - 2\varepsilon\zeta(\tau_1))u_0 + 2\varepsilon\lambda_0(u_0 - u_1) + 8\varepsilon\alpha u_0^3 = 2\varepsilon F \sin\tau_0,$$

where $\zeta(\tau_1) = s + b\tau_1$ denotes frequency detuning.

The asymptotic analysis of Eqs. (3) is performed with the help of the method of multiple scales [18]. To employ this method, we introduce the complex amplitudes

$$Y_r = (v_r + iu_r)e^{-i\tau_0}. \quad (4)$$

and then transform system (3) into the equations

$$\frac{dY_1}{d\tau_0} = i\varepsilon\lambda_1(Y_1 - Y_0) + i\varepsilon G_1,$$
$$\frac{dY_0}{d\tau_0} = -i\varepsilon[\zeta(\tau_1) - 3\alpha|Y_0|^2]Y_0 + i\varepsilon[\lambda(Y_0 - Y_1) - F] + i\varepsilon G_0 \quad (5)$$

with initial conditions $Y_0(0) = Y_1(0) = 0$. The coefficients $G_0$ and $G_1$ include the sum of harmonics with the coefficients depending on $Y_0$, $Y_1$, and their complex conjugates but their explicit expressions are insignificant for further analysis.

It follows from Eqs. (5) that main terms in the multiple scales expansion of amplitudes (4) are independent of $\tau_0$. This implies the expansion in the form

$$Y_r(\tau_0, \tau_1, \varepsilon) = \varphi_r(\tau_1) + \varepsilon\varphi_r^{(1)}(\tau_0, \tau_1) + ... \quad (6)$$

In order to diminish the number of the system parameters, the variables and the parameters are redefined as follows:

$$\Lambda = (s/3\alpha)^{1/2}, f = F/s\Lambda, \beta = b/s^2, \psi_r = \varphi_r/\Lambda, \mu_r = \lambda_r/s, \quad (7)$$
$$\tau = s\tau_1, \zeta_0(\tau) = 1 + \beta\tau, \; r = 0, 1.$$

Inserting (6), (7) into (5) and then employing the multiple scales formalism [18], we exclude non-oscillating terms from the resulting equations to obtain the following equations for the slow variables $\psi_0(\tau)$ and $\psi_1(\tau)$:

$$\frac{d\psi_1}{d\tau} - i\mu_1(\psi_1 - \psi_0) = 0, \psi_1(0) = 0,$$
$$\frac{d\psi_0}{d\tau} - i\mu_0(\psi_0 - \psi_1) + i(\zeta_0(\tau) - |\psi_0|^2)\psi_0 = -if, \psi_0(0) = 0. \quad (8)$$

It follows from (8) that the nonlinear response $\psi_0(\tau)$ acts as an external excitation for the linear oscillator. The real-valued amplitudes and phases of oscillations are defined as $a_r = |\psi_r|$ and $\Delta_r = \arg(\psi_r)$, respectively. The detailed derivation of the leading-order equations for similar systems has been presented in [19, 20].

Now we show that in some special cases the equation of the excited oscillator can be solved independently. In analogy with [20], we consider an asymmetric system in which $m_1 = \varepsilon\delta m_0$, $\delta = O(1)$. In this case, $\mu_0 = \mu_1 c_1/c_0 = \mu_1 m_1/m_0 = \varepsilon\delta\mu_1$, and thus the term proportional to $\mu_0$ may be removed from (8) in the main approximation. The resulting truncated system is given by

$$\frac{d\psi_1^{(0)}}{d\tau} - i\mu_1\psi_1^{(0)} = -i\mu_1\psi_0^{(0)}, \psi_1^{(0)}(0) = 0,$$
$$\frac{d\psi_0^{(0)}}{d\tau} + i(\zeta_0(\tau) - |\psi_0^{(0)}|^2)\psi_0^{(0)} = -if, \psi_0^{(0)}(0) = 0. \quad (9)$$

The nonlinear equation in system (9) can be investigated separately. This implies that if the approximate solutions $\psi_r^{(0)}(\tau)$ are close to the exact solution $\psi_r(\tau)$ ($r = 0,1$), then the condition of the occurrence of AR in a single Duffing oscillator [16, 17] can be extended to the weakly coupled system. In analogy with the earlier developed procedure [20], the effect of weak coupling may be considered in subsequent iterations. It is important to note that, although the assumption $\mu_0/\mu_1 \ll 1$ simplifies the analysis, the qualitative features of the dynamical behavior hold true for a wide range of the parameters such that $\mu_0 < 1$ and $\mu_1 < 1$.

We recall main results of earlier works [16, 17] required for further analysis. It was shown that AR in the Duffing oscillator may occur at $f > f_1 = \sqrt{2/27}$, while the values $f < f_1$ corresponds to bounded oscillations at any rate $\beta$. In the domain $f > f_1$ the Duffing oscillator admits AR if $\beta < \beta^*$ and bounded oscillations (saturation) if $\beta > \beta^*$. The critical rate is defined as $\beta^* = [(f/f_1)^{2/3} - 1]/T^*$, where $\tau = T^*$ corresponds to the first minimum of the phase $\Delta_0(\tau)$ in the time-independent Duffing oscillator ($\beta = 0$) [17]. The value of $T^*$ was found both numerically and analytically.

In analogy with a single oscillator (e.g., [21]), the solution $\psi_0(\tau)$ is considered as small fast fluctuations $\tilde{\psi}_0(\tau)$ near the quasi-steady state $\bar{\psi}_0(\tau)$, that is, $\psi_0(\tau) = \bar{\psi}_0(\tau) + \tilde{\psi}_0(\tau)$, where $\bar{\psi}_0$ is calculated as a stationary point of the system with "frozen" detuning $\zeta_0$. Assuming $\mu_0 = O(\varepsilon)$, we obtain the following equation for $\bar{\psi}_0(\tau)$:

$$(\zeta_0 - |\bar{\psi}_0|^2)\bar{\psi}_0 = -f. \quad (10)$$

If $f \ll 1$, the backbone curve $\bar{a}_0 = |\bar{\psi}_0|$ satisfies the equation

$$\bar{a}_0 \approx \sqrt{\zeta_0} \to \sqrt{\beta\tau} \text{ as } \tau \to \infty, \quad (11)$$

the corresponding phase $\bar{\Delta}_0 = \arg\bar{\psi}_0 = 0$. By definition, equality (11) expresses a relationship between the amplitude and the frequency of free oscillations [18]. In analogy to the conservative case [22], the asymptotic approximation for fast fluctuations $\tilde{\psi}_0(\tau)$ can be computed by linearizing the nonlinear equation near the quasi-steady solution $\bar{\psi}_0(\tau)$ and then omitting the term proportional to $\mu_0$.

If the solution $\psi_0(\tau)$ is known, then the response $\psi_1(\tau)$ is directly calculated from (8) by formula

$$\psi_1(\tau) = -i\mu_1 \int_0^\tau e^{i\mu_1(\tau-s)}\psi_0(s)ds. \quad (12)$$

Since the effect of small fast fluctuations $\tilde{\psi}_0(\tau)$ on the value of integral (12) is negligible compared to the contribution of the slowly-varying function $\bar{\psi}_0$, the following approximation is valid:

$$\psi_1(\tau) \approx -i\mu_1 e^{-i\mu_1\tau}J(\tau), J(\tau) = \int_0^\tau e^{i\mu_1 s}\bar{\psi}_0(s)ds \quad (13)$$

where $\bar{\psi}_0(\tau) = \sqrt{1+\beta\tau}$. Integration by parts gives

$$J(s) = -i\mu_1^{-1}[e^{i\mu_1\tau}\bar{\psi}_0(\tau) - \bar{\psi}_0(0)] - \Phi(\tau),$$

$$\Phi(\tau) = \frac{\beta}{2}\int_0^\tau \frac{e^{i\mu_1 s}}{\sqrt{1+\beta s}}ds.$$

It is easy to deduce that $\Phi(\tau) = \sqrt{\beta}F(\tau)$, where $F(\tau)$ is the Fresnel-type integral bounded at any $\tau > 0$. Hence $\psi_1(\tau) = \bar{\psi}_1(\tau) + \tilde{\psi}_1(\tau) + O(\sqrt{\beta})$, where

$$\bar{\psi}_1(\tau) = \bar{\psi}_0(\tau), \tilde{\psi}_1(\tau) \approx -\bar{\psi}(0)e^{i\mu_1\tau}. \quad (14)$$

Note that the condition $\bar{\psi}_1(\tau) = \bar{\psi}_0(\tau)$ as well as equalities (10), (11) can be directly deduced from Eqs. (8) but the performed transformations provides the formal demonstration of the occurrence of AR in the linear oscillator.

The obtained theoretical results are illustrated in Fig. 1. Throughout this paper, the following parameters are used for numerical computations:

$$\beta = 0.05, \mu = 0.02, \mu_1 = 0.25; f = 0.34. \quad (15)$$

Since $\beta^* \approx 0.06$ at $f = 0.34$, the system with chosen parameters admits AR (see [17]).

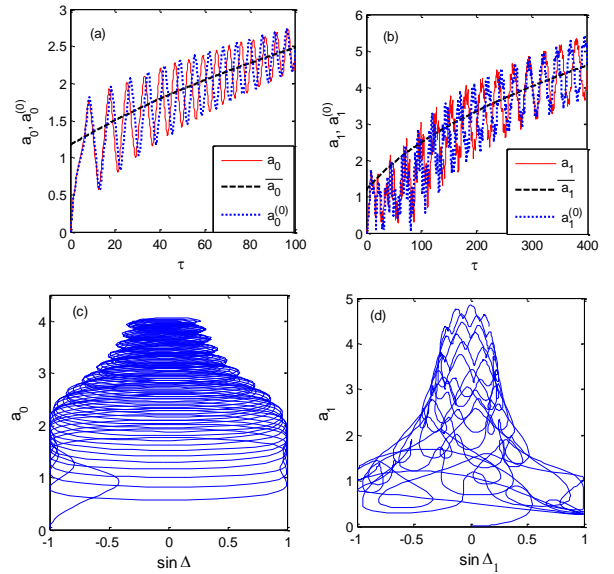

**Fig. 1.** Amplitudes and phases of oscillations

It is seen in Figs. 1(*a*) and 1(*b*) that the approximate (dotted lines) and exact (solid lines) amplitudes are close to each other. This implies that the conditions of the occurrence of AR in a single oscillator (9) can be extended to the weakly coupled system. Furthermore, fluctuations become negligible compared to the value of $\bar{a}_0(\tau)$ (Fig. 1(*a*)). This allows calculations of the asymptotic approximation for $\tilde{\psi}_0(\tau)$ by linearizing Eqs. (8) near the quasi-steady state $\bar{\psi}_0(\tau)$. Details are omitted for brevity.

Figure 1(*b*) demonstrates irregular oscillations of the linear oscillator at the early stage of motion but at later times forcing with an increasing amplitude dominates and motion is transformed into regular oscillations near the backbone curve $\bar{a}_1(\tau)$. In the system with coupling $\mu_1 = 0.25$ we obtain the period of oscillations $T_1 = 2\pi/\mu_1 \approx 25.12$ and the amplitude $\tilde{a}_1 = \bar{\psi}(0) \approx 1$; both these values are close to the corresponding parameters in Fig. 1(*b*). Figures 1(*c*) and 1(*d*) illustrate phase locking in AR oscillations

As shown above, the resonance state of the system is efficiently controlled via the change of linear stiffness in the Duffing oscillator, and a required state can be maintained by terminating the change of the parameter at the prescribed energy level. Since the amplitudes (and energy) of oscillations can be represented as small fast fluctuations near the monotonically increasing backbone curves, it is convenient to evaluate the terminal time $T^*$ at which the backbone curve $\bar{a}_1(\tau)$ achieves the required value $\bar{a}_1(T^*) = \varepsilon_1$. It follows from (11) that $T^*$ can be evaluated by formula $(1 + \beta T^*) = \varepsilon_1^2$.

Figure 2 depicts the amplitudes of oscillations for the system with parameters (15) and the terminal time $T^* = 200$. It is seen that at $\tau > T^*$ AR turns into oscillations with almost constant amplitudes, and the theoretical value $\bar{a}_1(T^*) \approx 3.3$ is close to the result presented in Fig. 2

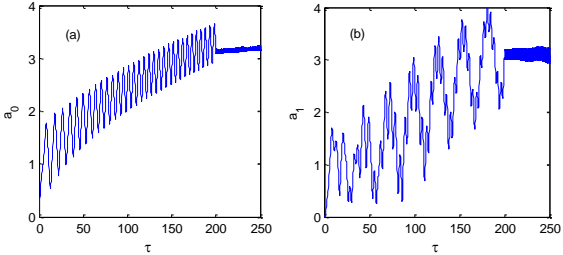

**Fig. 2.** Transitions from AR to oscillations with the prescribed energy in the nonlinear (*a*) and linear (*b*) oscillators

**Energy localization in the system of the second type.** - In this section we briefly analyse AR in the system with constant parameters driven by an external force with the slowly changing frequency. The equations of motion are reduced the form similar to (3):

$$\frac{d^2 u_0}{d\tau_0^2} + u + 2\varepsilon\lambda(u_0 - u_1) + 8\varepsilon\alpha u_0^3 = 2\varepsilon F \sin(\tau_0 + \theta(\tau_1)), \quad (16)$$

$$\frac{d^2 u_1}{d\tau_0^2} + u_1 + 2\varepsilon\lambda_1(u_1 - u_0) = 0,$$

where $d\theta/d\tau_1 = \zeta(\tau_1)$, $\tau_1 = \varepsilon\tau_0$. Transformations (4) - (7) and the change of variables $\phi_r = \psi_r e^{-i\theta_0(\tau)}$, $\tau = s\tau_1$, entail the following dimensionless equations for the slow complex amplitudes $\phi_r(\tau)$:

$$\frac{d\phi_1}{d\tau} + i\zeta_0(\tau)\phi_1 - i\mu_1(\phi_1 - \phi_0) = 0, \phi_1(0) = 0,$$

$$\frac{d\phi_0}{d\tau} - i\mu_0(\phi_0 - \phi_1) + i(\zeta_0(\tau) - |\phi_0|^2)\phi = -if, \phi_0(0) = 0. \quad (17)$$

It is important to note that system (17) also has the constant right-hand side but the time-dependent coefficient $\zeta_0(\tau)$ is now involved in both equations.

As in the previous section, the response $\phi_0(\tau)$ of the nonlinear oscillator is presented as the sum $\phi_0 = \bar{\phi}_0 + \tilde{\phi}_0$, where $\bar{\phi}_0(\tau)$ and $\tilde{\phi}_0(\tau)$ denote the quasi-steady state and small fast fluctuations near this state, respectively. Assuming $\mu_0 \ll \mu_1$, we find that the state $\bar{\phi}_0(\tau)$ and the backbone curve satisfies the equations similar to (10) and (11), respectively. Fast fluctuations $\tilde{\phi}_0(\tau)$ can be calculated by linearizing Eqs. (17) and disregarding the terms proportional to $\mu_0$.

After calculating the nonlinear response $\phi_0(\tau)$, the response of the linear oscillator $\phi_1(\tau)$ can be directly found from Eq. (17). Ignoring the effect of small fast fluctuations we obtain after simple transformations

$$\phi_1(\tau) \approx -i\frac{\mu_1}{2\beta}e^{-iS(\tau)/2\beta}K(\tau), \ S(\tau) = (1+\beta\tau)^2,$$

$$K(\tau) = K_0(\tau) - K_0(1), \ K_0(\tau) = \int_0^{S(\tau)} e^{iz/2\beta}z^{-1/4}dz. \quad (18)$$

Although the expression for $K_0(\tau)$ cannot be analytically found in the closed form, the limiting value $K_0(\infty)$ can be explicitly evaluated, and equals $K_0(\infty) = (2\beta)^{4/3}\Gamma(3/4)e^{3i\pi/8}$, where $\Gamma$ is the gamma function [23]. Hence,

$$a_1(\tau) = |\phi_1(\tau)| \to \mu_1(2\beta)^{1/3}\Gamma(3/4), \ \tau \to \infty. \quad (19)$$

Formula (19) indicates that AR in the nonlinear oscillator is unable to generate oscillations with permanently growing energy in the attached oscillator but it suffices to produce linear oscillations with bounded amplitude. The substitution of parameters (15) into (19) defines the limiting amplitude $a_{1\infty} = \lim_{\tau\to\infty} a(\tau) \approx 0.1$.

It is seen from Fig. 3 that the amplitude of nonlinear oscillations is very close to its analogue presented in Fig. 1(*a*) but the behavior of the linear oscillator radically differs from oscillations with growing energy presented Fig. 1(*b*). The shape of the amplitude $a_1(\tau)$ in Fig. 3 is similar to the resonance curve with a noticeable resonance peak at the initial stage of motion where the effect of the time-dependent detuning is negligible, but then it turns into small oscillations with the limiting amplitude tending to $a_{1\infty}$.

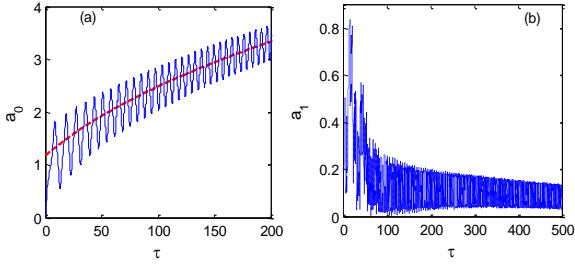

**Fig. 3.** Amplitudes of oscillations of the nonlinear (*a*) and linear (*b*) oscillators with parameters (15); dashed line in plot (*a*) denotes the backbone curve $\bar{a}_0$.

A key conclusion from the obtained results is that in the system of the second type the energy transferred from the nonlinear oscillator is insufficient to produce oscillations with growing energy in the attached linear oscillator. The different dynamical behavior can be considered as a consequence of different resonance properties of the systems. In the system with a constant frequency of external forcing both oscillators are captured in resonance. If the forcing frequency slowly increases and the parameters of the system remain constant, resonance in the nonlinear oscillator is still sustained by increasing amplitude, while the frequency of the linear oscillator falls into the domain beyond the resonance. This conclusion is by no means trivial, as the linear oscillator is actually driven by the coupling response with permanently increasing amplitude.

**Saturation in the 2D system.** - In this section we briefly discuss the occurrence of bounded oscillations in the coupled system of the first type. It was demonstrated in earlier works [16, 17] that the transition from AR to bounded oscillations in a single Duffing oscillator occurs at rate $\beta > \beta^*$. Figure 4 demonstrates a similar effect in the coupled oscillators with parameters (15) and detuning rate $\beta = 0.065 > \beta^* = 0.06$. The existence of bounded oscillations in the nonlinear oscillator at $\beta > \beta^*$ may be predicted referring to the approximation of the solution $\psi_0(\tau)$, which characterizes the dynamics of the Duffing oscillator in the coupled system, by the function $\psi^{(0)}(\tau)$ describing the dynamics of the uncoupled nonlinear oscillator. The bounded solution for the linear oscillator can be directly obtained from the equation similar to (13).

We recall that the transition from AR to saturation in a single Duffing oscillator is of the same nature as the transition from large to small oscillations in the system with constant parameters and occurs due to the destruction of the LPT of large oscillations (see [16, 17] for detail). Figure 4 demonstrates an analogous process in the coupled system. Note that the limit values $\bar{a}_0 = \lim_{\tau \to \infty} a_0(\tau)$ and $\bar{a}_1 = \lim_{\tau \to \infty} a_1(\tau)$ as $\tau \to \infty$ (dashed lines in Fig. 4) cannot be considered as the quasi-steady states of system (8) at "frozen" $\zeta_0$.

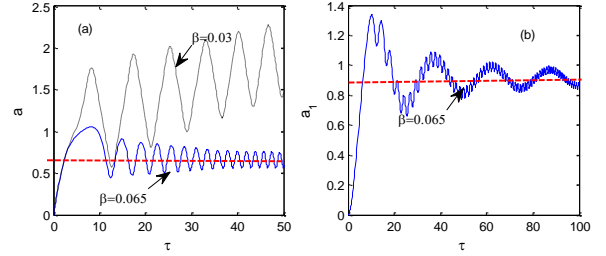

**Fig. 4.** Amplitudes of saturated oscillations; dashed lines depict the limiting levels $\bar{a}_0$ and $\bar{a}_1$; dotted line corresponds to autoresonance at $\beta = 0.03$.

The regime of saturation in the nonlinear oscillator can be approximately described with the help of the iteration procedure developed in [21]. In the first step, the initial iterations $\Psi_0$, $\Psi_1$ are computed as the solutions of the linear system

$$\frac{d\Psi_1}{d\tau} - i\mu_1(\Psi_1 - \Psi_0) = 0, \Psi_1(0) = 0, \qquad (20)$$
$$\frac{d\Psi_0}{d\tau} + i\zeta_0(\tau)\Psi_0 = -if, \Psi_0(0) = 0.$$

It is easy to obtain from the second equation that the function $\Psi_0$ is expressed through the Fresnel-type integrals. However, the substitution of the Fresnel-type solution $\Psi_0(\tau)$ into the equation for $\Psi_1$ yields a complicated combination of the Fresnel integrals and exponential functions, which is very difficult for the direct analysis. So, although explicit closed-form approximations are formally available, the amplitudes $a_0(\tau)$ and $a_1(\tau)$ as well as the limit values $\bar{a}_0$ and $\bar{a}_1$ should be obtained numerically.

Subsequent iterations can be found from the equations

$$\frac{d\Psi_{1,n}}{d\tau} - i\mu_1(\Psi_{1,n} - \Psi_{0,n}) = 0, \Psi_{1,n}(0) = 0,$$
$$\frac{d\Psi_{0,n}}{d\tau} + i\zeta_0(\tau)\Psi_{0,n} = -if + |\Psi_{0,n-1}|^2 \Psi_{0,n-1} + \qquad (21)$$
$$+ i\mu_0(\Psi_{0,n-1} - \Psi_{1,n-1}), \Psi_{0,n}(0) = 0, n \geq 1.$$

**Conclusions.** – It was shown in early works on particle acceleration that autoresonance (AR) could potentially serve as a tool for excitation and control of the required high-energy regime in a single oscillator. This method of excitation was further employed in various fields of the applied physics and engineering. However, the behavior of coupled oscillators may drastically differ from the dynamics of a single oscillator. In particular, the capture into resonance may not exist or AR in one part of the system may be insufficient to enhance the response of other oscillators. We have investigated these effects in the 2D system of coupled linear and nonlinear oscillators with different types of perturbations: (1) a periodic force with constant (resonance) frequency is applied to the nonlinear (Duffing) oscillator with slowly time-decreasing linear stiffness; (2) the nonlinear oscillator with constant parameters is excited by a force with slowly increasing frequency. In both cases, the parameters of the linear oscillator and linear coupling are constant, and the system is initially engaged in resonance. It has been shown that in the system of the first type AR occurs in both oscillators but in the system of the second type energy transfer from the nonlinear oscillator is insufficient to excite high-energy motion in the attached oscillator. In the systems of this type, AR in the nonlinear oscillator is accompanied by small oscillations of the attached oscillator. This implies that energy transfer from the nonlinear oscillator may generate the high-energy regime in the linear oscillator only in the system of the first type, while in the system of the second type energy remains localized on the excited nonlinear oscillator. To the best of the authors' knowledge, this effect has not been reported thus far in the literature.

**Acknowledgement.** - The authors acknowledge support for this work from the Russian Foundation for Basic Research through the RFBR grant 14-01-00284.